\documentclass[conference]{IEEEtran}
\IEEEoverridecommandlockouts

\usepackage{cite}
\usepackage{multirow}
\usepackage{subcaption}
\usepackage{amsmath,amssymb,amsfonts}
\usepackage{algpseudocode}
\usepackage{algorithm}
\usepackage{tabularx}
\usepackage{booktabs}
\usepackage{array}
\usepackage{graphicx}
\usepackage{textcomp}
\usepackage{multirow}
\usepackage{xcolor}
\usepackage{balance}
\usepackage{comment}
\def\BibTeX{{\rm B\kern-.05em{\sc i\kern-.025em b}\kern-.08em
    T\kern-.1667em\lower.7ex\hbox{E}\kern-.125emX}}
\begin{document}

\title{Modeling Quantum Federated Autoencoder for Anomaly Detection in IoT Networks}

\author{
\IEEEauthorblockN{Devashish Chaudhary\textsuperscript{*}, 
Sutharshan Rajasegarar\textsuperscript{†}, 
Shiva Raj Pokhrel\textsuperscript{‡}}
\IEEEauthorblockA{\textit{School of Information Technology, Deakin University, Geelong, Australia} \\
\textsuperscript{*}s224281473@deakin.edu.au,
\textsuperscript{†}srajas@deakin.edu.au,
\textsuperscript{‡}shiva.pokhrel@deakin.edu.au}
}

\maketitle

\begin{abstract}
We propose a Quantum Federated Autoencoder for Anomaly Detection, a framework that leverages quantum federated learning for efficient, secure, and distributed processing in IoT networks. By harnessing quantum autoencoders for high-dimensional feature representation and federated learning for decentralized model training, the approach transforms localized learning on edge devices without requiring transmission of raw data, thereby preserving privacy and minimizing communication overhead. The model leverages quantum advantage in pattern recognition to enhance detection sensitivity, particularly in complex and dynamic IoT network traffic. Experiments on a real-world IoT dataset show that the proposed method delivers anomaly detection accuracy and robustness comparable to centralized approaches, while ensuring data privacy.
\end{abstract}


\section{Introduction}

With the recent accelerated growth of interconnected devices, securing the network from attacks and timely detection of emerging anomalies have become increasingly challenging ~\cite{singh2025empowering}. Conventional approaches require transmission of raw data to a centralized server for model training, which introduces severe privacy risks and creates a single point of failure; any compromise of the server might expose the entire dataset. Federated Learning (FL)~\cite{nguyen2021federated, chaudhary2025towards} mitigates this issue by training models locally on devices and sharing only the model parameters with a server, which then aggregates updates to form a global model~\cite{nanayakkara2024understanding}.

Recent advances in quantum computing provide new capabilities for machine learning by exploiting superposition and entanglement to process complex data more efficiently. Integrating these capabilities into FL yields Quantum Federated Learning (QFL)~\cite{10930694}, where quantum models are trained locally, and only parameter updates are communicated for aggregation. QFL holds significant promise for enhancing both performance and security in large-scale distributed networks, particularly for anomaly detection.
\begin{figure}[t]
    \centering
    \includegraphics[width=0.69\linewidth]{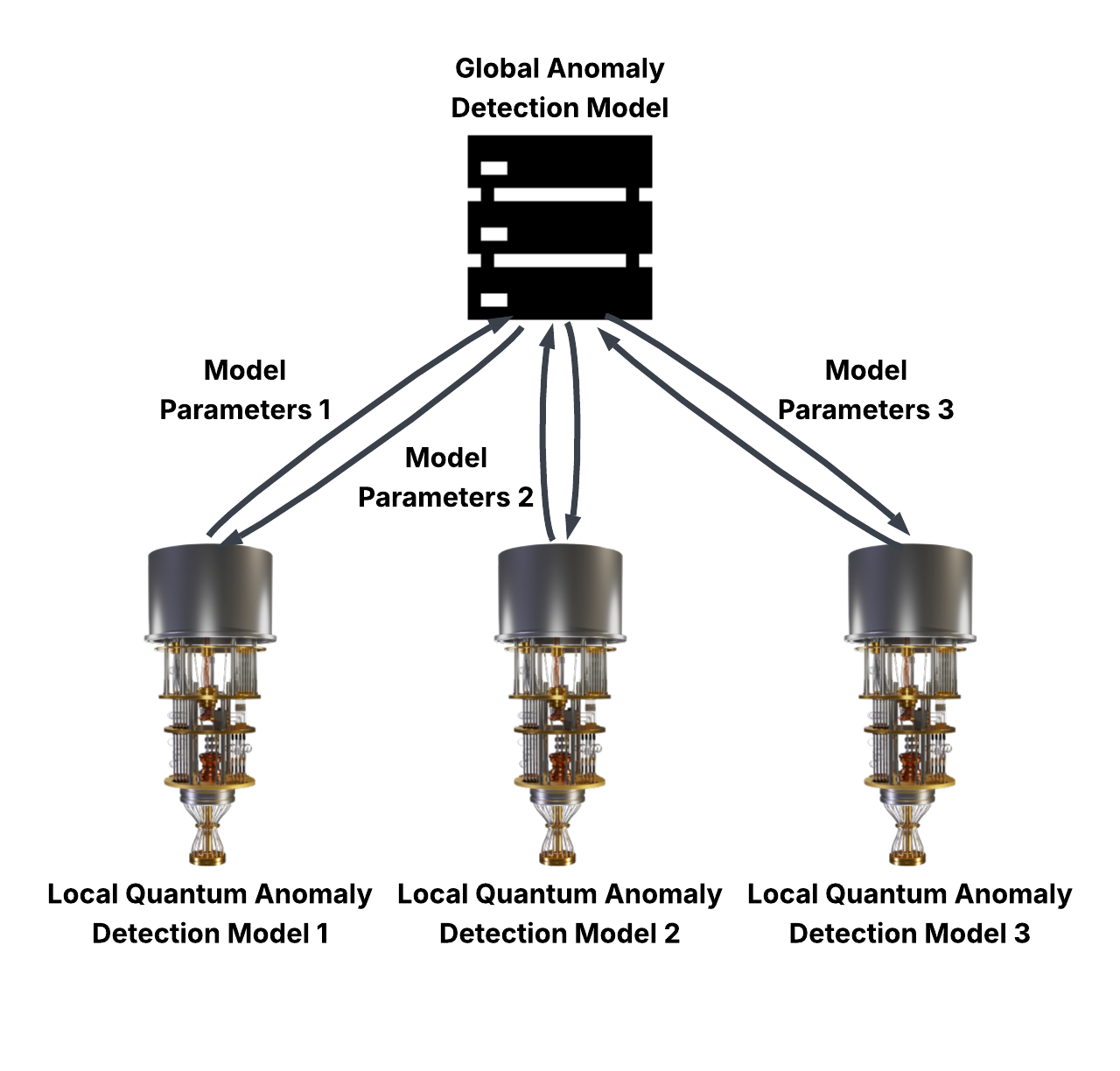}
    \caption{Quantum Federated Autoencoder Framework for Anomaly Detection.}
    \label{fig:fl}
\end{figure}



The primary contributions of this work are summarized below:

\begin{itemize}
\item We construct a fully operational hierarchical IoT network using Raspberry~Pi~3B+ devices and XBee transceivers, enabling the generation of realistic, multilevel traffic patterns. This testbed provides high-fidelity data streams from which salient features for anomaly detection are systematically extracted.

\item We develop a new QFL architecture (Fig.~\ref{fig:fl}) that enables the local training of quantum autoencoders on edge devices and supports both hierarchical and canonical FedAvg aggregation strategies. The framework is rigorously benchmarked against an equivalent centralized quantum-training baseline.

\item Comprehensive experiments across heterogeneous IoT devices show that the proposed QFL framework preserves data privacy without degrading model utility, achieving detection performance statistically indistinguishable from fully centralized quantum training.
\end{itemize}

To the best of our knowledge, this work constitutes the first demonstration of a quantum autoencoder adapted for anomaly detection and deployed within a fully operational QFL setting.

\section{Related work}

Classical deep learning approaches have been extensively used for anomaly detection in networks~\cite{wu2020network}, leveraging both supervised and unsupervised models. Many works have combined autoencoders with temporal architectures to capture dynamic traffic behavior. For instance, a PSO-Autoencoder-LSTM framework~\cite{narmadha2025improved} integrates autoencoders for feature extraction with LSTMs for temporal dependency modeling, while Particle Swarm Optimization (PSO) tunes hyperparameters to improve detection accuracy and resilience against class imbalance in intrusion detection tasks. Despite their effectiveness, these approaches~\cite{wu2020network, narmadha2025improved} remain inherently centralized, requiring raw data aggregation at a server, which exposes sensitive traffic to privacy and security risks.

FL addresses these limitations by enabling decentralized training without exposing raw data. Instead, clients train models locally and transmit only parameter updates to a central server for aggregation. Recent research has explored FL-based anomaly detection in network data. For example, Fed-ANIDS~\cite{idrissi2023fed} deploys simple, variational, and adversarial autoencoders across distributed clients, employing FedAvg or FedProx for aggregation. Clients share model updates derived from local benign traffic, while the global model uses reconstruction error thresholds to detect anomalies. Such approaches effectively preserve privacy while maintaining strong intrusion detection performance.

Nevertheless, IoT traffic often exhibits highly complex, nonlinear and high-dimensional patterns that classical models struggle to capture efficiently. QML offers a promising alternative by exploiting superposition and entanglement to model such complexity more effectively. Hdaib et al.~\cite{hdaib2024quantum} demonstrated the potential of autoencoders for anomaly detection, benchmarking their integration with quantum one-class SVM, quantum k-nearest neighbor, and quantum random forest classifiers. Their results showed that an autoencoder combined with quantum kNN achieved superior performance, underscoring the advantage of quantum models. Nevertheless, existing QML methods are centralized and thus suffer from the same privacy and scalability issues as classical deep learning. To date, no study has explored a federated paradigm that combines the privacy-preserving benefits of FL with the representational power of quantum autoencoders.

\section{Proposed Quantum Federated Autoencoder}
\begin{figure}[t]
    \centering
    \includegraphics[width=1.00\linewidth]{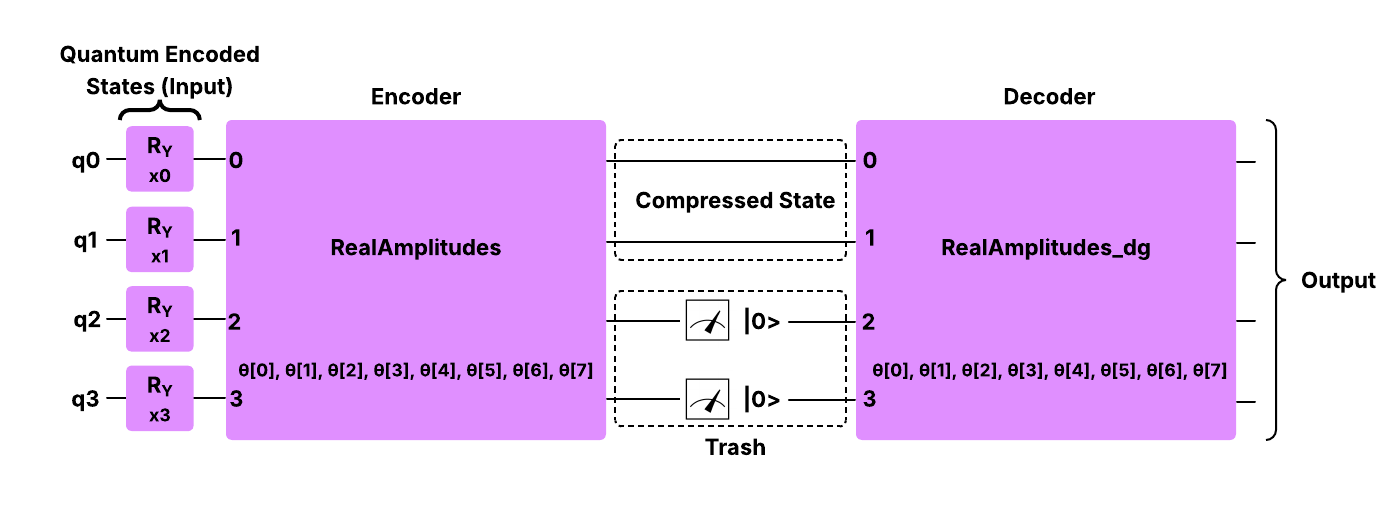}
    \caption{Quantum Autoencoder with 4 qubits: 3 latent, 1 trash.}
    \label{fig:qae}
\end{figure}

Fig.~\ref{fig:qae} shows a minimal 4-qubit Quantum autoencoder (QAE) for illustration. The experiments use a 10-qubit model (8 latent, 2 trash) as described in Section~\ref{tc}. Different from classical autoencoders, which reduce the dimensionality of input data by learning a compact representation that can be decoded to reconstruct the original input, QAEs extend to the quantum domain, efficiently compressing quantum states stored on $n$ qubits into a smaller set of $m < n$ qubits. 
\begin{algorithm}[!h]
\caption{Federated Quantum Autoencoder Training}
\label{alg: fed}
\begin{algorithmic}[1]
\Procedure{TrainQAE}{$X$, $d$, $n$, $R$, $F$, $I$}
    \State \textbf{Input:} Network traffic data $X$, PCA components $d=10$, qubits $n=10$, routers $R=3$, federated rounds $F=5$, local iterations $I=50$
    \State \textbf{Output:} Trained global QAE parameters $\boldsymbol{\theta}_{\text{global}}$
    
    \State Perform PCA on $X$ to reduce dimensionality to $d$ components
    \For{$r = 1$ to $R$}
        \State Encode PCA features into $n$ qubits using angle encoding ($R_y$ rotations)
        \State Initialize parameterized quantum circuit (RealAmplitude) as encoder
    \EndFor

    \For{$t = 1$ to $F$} \Comment{Federated rounds}
        \For{$r = 1$ to $R$} \Comment{Routers process in parallel}
            \For{$i = 1$ to $I$} \Comment{Local training iterations}
                \State Forward pass through QAE to compute probabilities
                \State Compute loss $L(\boldsymbol{\theta})$: Equation (1) 
                
                
                \State Update parameters $\boldsymbol{\theta}_r$ to minimize $L(\boldsymbol{\theta})$
            \EndFor
        \EndFor
        \State Routers send local parameters to coordinator
        \State Coordinator performs hierarchical federated averaging (\textit{hierarchical FL}) or \textit{FedAvg} to obtain $\boldsymbol{\theta}_{\text{global}}$
        \State Coordinator sends updated global parameters back to all routers
    \EndFor
    \State \Return $\boldsymbol{\theta}_{\text{global}}$
\EndProcedure
\end{algorithmic}
\end{algorithm}

The proposed framework leverages QFL to enable distributed training of QAEs across local devices while preserving data privacy. In the proposed QFL, each device trains a QAE locally and transmits only model parameters to a central server for aggregation. Aggregation can follow a standard \textit{FedAvg} or a \textit{hierarchical FL} protocol. In the \textit{hierarchical FL}, updates from child nodes are first combined at parent nodes before progressively forming a global model (See Fig.~\ref{fig:fl}, Algorithm~\ref{alg: fed}).
\begin{table*}[ht]
  \centering
  \caption{Attack scenarios: C – Coordinator, A – Attacker, Rx – Router, Ex – Edge/Leaf device.}
  \resizebox{\textwidth}{!}{%
    \begin{tabular}{r|l|l|l|l|l|l|l|l|l|l|l|l|l|l}
    \hline
    \multicolumn{5}{c|}{\textbf{Scenario I}} & \multicolumn{3}{c|}{\textbf{Scenario II}} & \multicolumn{7}{c}{\textbf{Scenario III}} \\
    \hline
    \multicolumn{1}{l|}{\textbf{Normal}} & E1 $\rightarrow$ R1 & E2 $\rightarrow$ R1 & E3 $\rightarrow$ R3 & E4 $\rightarrow$ R2 & \multicolumn{1}{l|}{R1 $\rightarrow$ C} & \multicolumn{1}{l|}{R2 $\rightarrow$ C} & \multicolumn{1}{l|}{R3 $\rightarrow$ R2} & \multicolumn{1}{l|}{E1 $\rightarrow$ R1} & \multicolumn{1}{l|}{E2 $\rightarrow$ R1} & \multicolumn{1}{l|}{E3 $\rightarrow$ R3} & \multicolumn{1}{l|}{E4 $\rightarrow$ R2} & \multicolumn{1}{l|}{R1 $\rightarrow$ C} & \multicolumn{1}{l|}{R2 $\rightarrow$ C} & \multicolumn{1}{l}{R3 $\rightarrow$ C} \\
    \hline
    \textbf{Attack} & E1 $\rightarrow$ R2 & E2 $\rightarrow$ R2 & E3 $\rightarrow$ R1 & E4 $\rightarrow$ R1 & \multicolumn{1}{l|}{R1 $\rightarrow$ R2} & \multicolumn{1}{l|}{R2 $\rightarrow$ R1} & \multicolumn{1}{l|}{R3 $\rightarrow$ R1} & \multicolumn{1}{l|}{E1 $\rightarrow$ A} & \multicolumn{1}{l|}{E2 $\rightarrow$ A} & \multicolumn{1}{l|}{E3 $\rightarrow$ A} & \multicolumn{1}{l|}{E4 $\rightarrow$ A} & \multicolumn{1}{l|}{R1 $\rightarrow$ A} & \multicolumn{1}{l|}{R2 $\rightarrow$ A} & \multicolumn{1}{l}{R3 $\rightarrow$ A} \\
    & E1 $\rightarrow$ R3 & E2 $\rightarrow$ R3 & E3 $\rightarrow$ R2 & E4 $\rightarrow$ R3 & \multicolumn{1}{l|}{R1 $\rightarrow$ R3} & & \multicolumn{1}{l|}{R3 $\rightarrow$ C} & & & & & & & \\
    & E1 $\rightarrow$ C & E2 $\rightarrow$ C & E3 $\rightarrow$ C & E4 $\rightarrow$ C & & & & & & & & & & \\
    \hline
    \end{tabular}%
  }
  \label{tab:AttScenarios}%
\end{table*}

A QAE compresses quantum states by mapping an input \(|\psi\rangle\) into a tensor product of compressed and ``trash'' qubits:  
\[
U |\psi\rangle = |\phi\rangle_C \otimes |0\rangle_T,
\]  
where \(|\phi\rangle_C\) retains relevant information and \(|0\rangle_T\) is the fixed state for the trash qubits. Reconstruction is achieved by applying the inverse unitary:  
\[
U^\dagger (|\phi\rangle_C \otimes |0\rangle_T) = |\psi\rangle.
\]  

Unlike classical autoencoders, QAEs cannot discard qubits, so the network explicitly minimizes residual information in trash qubits. For a system with \(n\) qubits, where the last \(k\) qubits form the trash subsystem, we define a cost function as the total probability of measuring ``bad'' states in the trash qubits (all basis states except \(|0\rangle_T\)):  
\begin{equation}
    L(\boldsymbol{\theta}) = \frac{1}{N} \sum_{i=1}^{N} \sum_{b \in \text{bad states}} 
\Pr[\text{trash qubits in state } b \mid X_{\text{train}}^{(i)}, \boldsymbol{\theta}],
\end{equation}
with \(\Pr[\cdot]\) obtained from the parameterized quantum circuit and \(N\) the batch size. Minimizing this loss ensures that the compressed qubits retain meaningful information while the trash qubits converge to \(|0\rangle_T\).

By adapting QAE into the QFL framework, our approach achieves privacy-preserving, distributed quantum compression and anomaly detection across IoT devices, combining the representational power of quantum models with the scalability of FL.

\section{Experimental Setup}
\begin{figure}[t]
    \centering
    \includegraphics[width=0.8\linewidth]{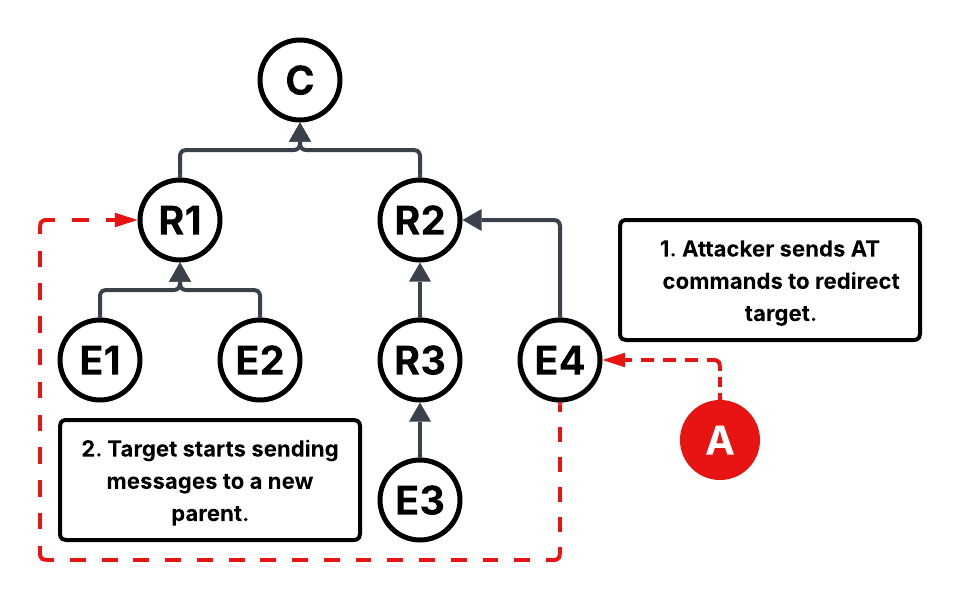}
    \caption{Network topology with Coordinator (C), Routers (R), Edge nodes (E) and Attack node (A).}
    \label{fig:topology}
\end{figure}

We have implemented a real-world IoT testbed consisting of Raspberry Pi 3B+ devices equipped with XBee S2C ZigBee modules for wireless communication \cite{Devashish25In}. The network was modelled hierarchically, as illustrated in Fig.~\ref{fig:topology}. 
One node was designated as an attacker to launch security attacks and generate both benign and malicious network traffic. Each edge device (E) sends data packets with timestamps. Intermediate routers (R) record reception timestamps before forwarding packets to the next hop, while the coordinator (C) maintains a complete log of packet paths and timings.  

Initially, network traffic logs were recorded under normal operating conditions. Relevant features were extracted for each one-minute time window, including mean and first-hop delay, delay quartiles, Shannon entropy, per-type and overall communication counts, and average hops per communication. Subsequently, redirection attacks were executed by exploiting ZigBee attention (AT) commands, generating malicious traffic for evaluation. Normal traffic was collected for 5 hours for training, with an additional 1 hour for validation. Each attack session included 20 minutes of normal traffic, 5 minutes of attack traffic, and 10 minutes of normal traffic, in a sequence. Three scenarios were created to generate attacks in the network, as shown in Table~\ref{tab:AttScenarios}.

Features were computed locally at each router for federated learning, whereas centralized training combined and shuffled data from all routers (R1, R2, R3). Preprocessing included MinMax scaling and principal component analysis (PCA), fitted on training data and applied to validation and test sets. For federated training, each router’s training set was partitioned into 5 subsets corresponding to 5 FL rounds.  

\subsection{Training Configuration}
\label{tc}

Training was conducted in Python 3.11.7 using Qiskit on a quantum simulator. The QAE utilized 10 qubits (8 latent, 2 trash), Ry rotations for feature transformation, and a RealAmplitude circuit as the parameterized encoder. Optimization employed COBYLA with 50 iterations for FL experiments and 100 iterations for centralized training. Federated learning was conducted over 5 rounds across 3 clients, with MinMax scaling and PCA for reducing 31 original features to 10 principal components. Experiments were conducted on an Apple MacBook Pro (M2, 8-core CPU, 10-core GPU).  

\subsection{Thresholding for Anomaly Detection}
Anomaly detection is performed using a threshold ($\tau$) computed from the reconstruction fidelity on the validation set: $\tau = \mu - 4\sigma$,
where \(\mu\) and \(\sigma\) are the mean and standard deviation of fidelities. Test samples with fidelity below \(\tau\) are classified as anomalies, providing a statistically robust cutoff that balances sensitivity and variability in reconstruction performance.

\section{Evaluation: Results and Discussion}

We compared two different types of FL methods with a centralized approach for anomaly detection in IoT networks.

\begin{table}[t]
\centering
\caption{Performance metrics across routers for FL methods}
\label{tab:performance_metrics}
\begin{tabular}{c|c|c|c|c|c}
\hline
\textbf{Routers} & \textbf{Method} & \textbf{Accuracy} & 
\textbf{Precision} &
\textbf{Recall} & \textbf{F1} \\ \hline
\multirow{3}{*}{R1} 
    & Hierarchical FL & 0.91 & 0.94 & 0.91 & 0.92 \\ 
    & FedAvg FL & 0.93 & 0.94 & 0.93 & 0.93 \\ 
    & Centralized & 0.88 & 0.93 & 0.88 & 0.90 \\ \hline
\multirow{3}{*}{R2} 
    & Hierarchical FL & 0.91 & 0.91 & 0.91 & 0.91 \\ 
    & FedAvg FL & 0.91 & 0.91 & 0.91 & 0.91 \\ 
    & Centralized & 0.91 & 0.91 & 0.91 & 0.91 \\ \hline
\multirow{3}{*}{R3} 
    & Hierarchical FL & 0.99 & 0.99 & 0.99 & 0.99 \\ 
    & FedAvg FL & 0.94 & 0.96 & 0.94 & 0.95 \\ 
    & Centralized & 0.99 & 0.99 & 0.99 & 0.99 \\ \hline
\end{tabular}
\end{table}

\begin{figure*}
    \centering
    \begin{subfigure}{1.00\linewidth}
        \centering
        \includegraphics[width=0.798\linewidth]{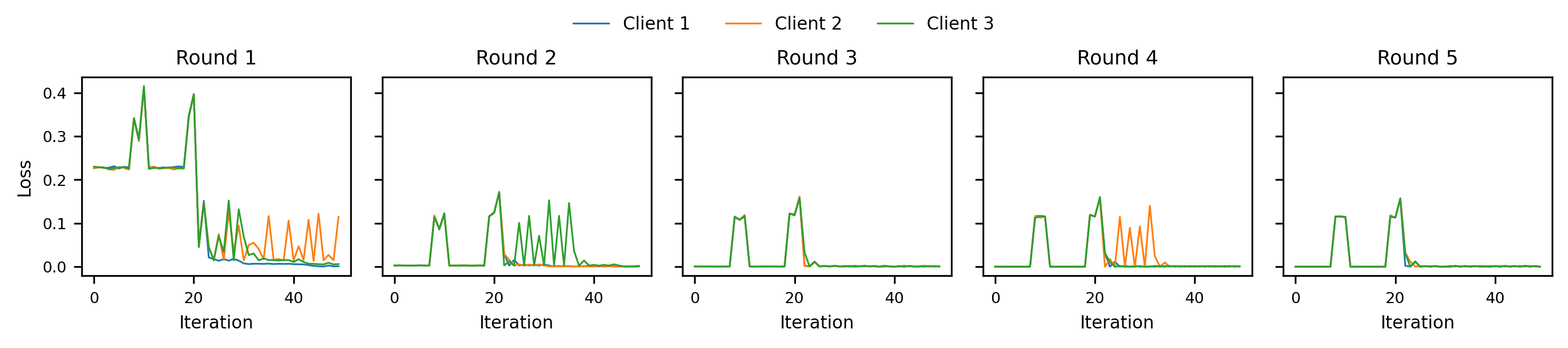}
        \caption{Hierarchical FL}
        \label{fig:Hierarchical}
    \end{subfigure}
    \hfill
    \begin{subfigure}{1.00\linewidth}
        \centering
        \includegraphics[width=0.798\linewidth]{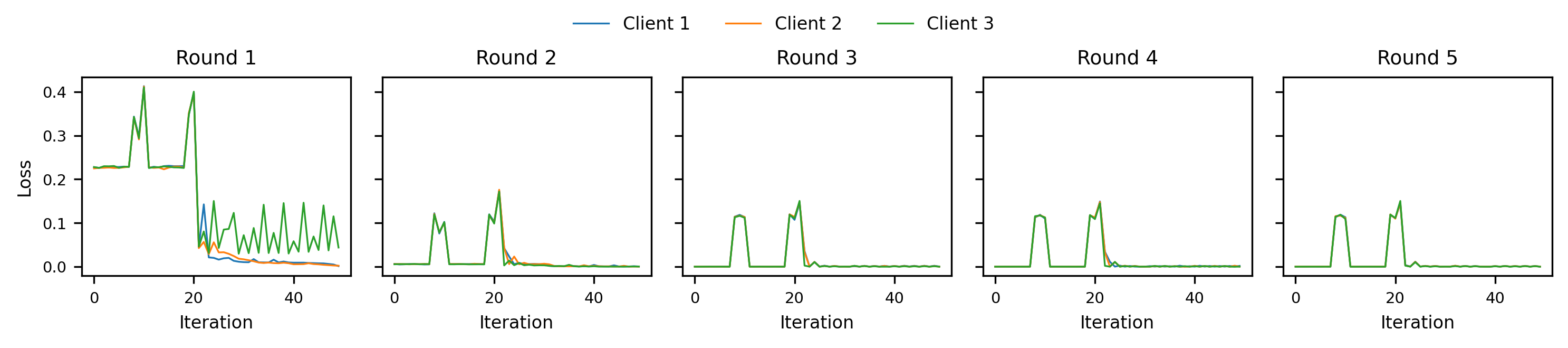}
        \caption{Standard FedAvg FL}
        \label{fig:Standard}
    \end{subfigure}
    \caption{Comparison of FL methods for 3 devices over 5 rounds: (a) Hierarchical FL and (b) Standard FedAvg FL.}
    \label{fig:FL_comparison}
\end{figure*}

Table~\ref{tab:performance_metrics} presents the performance metrics across the three routers for different learning methods. Overall, federated learning approaches deliver performance on par with or exceeding centralized training, demonstrating that the privacy-preserving distributed learning can maintain high anomaly detection effectiveness without requiring raw data aggregation.

Examining individual routers provides further insight into the strengths of hierarchical QFL. For Router 1, standard FedAvg slightly outperforms hierarchical QFL, while the centralized model shows marginally lower accuracy and F1-score, suggesting that uniform local patterns allow standard aggregation to perform well. For Router 2, all methods perform equivalently, indicating a simple or homogeneous data distribution across clients. In contrast, Router 3 highlights the advantage of hierarchical aggregation: both hierarchical FL and centralized training achieve near-perfect scores, whereas standard FedAvg falls slightly short. This demonstrates that hierarchical FL more effectively captures local variations and inter-node dependencies, which is particularly important in heterogeneous or non-uniform IoT networks. Collectively, these findings justify the adoption of hierarchical federated learning as a robust, privacy-preserving approach capable of matching or surpassing centralized methods.

Figures~\ref{fig:Hierarchical} and \ref{fig:Standard} show the evolution of training loss over iterations. Initial spikes in the first round reflect the model adapting to diverse local data distributions. In subsequent rounds, the loss begins to decrease and converges more smoothly, indicating that the models progressively internalize the underlying patterns across clients. This trend confirms that both standard and hierarchical QFL enable stable and effective distributed learning, with hierarchical aggregation providing additional resilience against local data heterogeneity.

Table~\ref{tab:performance_metrics} reveals that the federated approaches achieve accuracy and F1-scores comparable to the centralized baseline, demonstrating that privacy-preserving training does not compromise effectiveness. Router-specific outcomes further validate data heterogeneity: while FedAvg slightly outperforms in Router~1, hierarchical FL excels in Router~3, confirming the benefit of topology-aware aggregation. Figures~\ref{fig:Hierarchical} and~\ref{fig:Standard} show a smooth convergence of training loss across rounds, evidencing the stability of federated optimization. 

\section{Conclusion}
We evaluated anomaly detection using quantum autoencoders in a federated learning framework with both hierarchical and standard aggregation. Using data collected from a real Raspberry Pi 3B+ testbed, we showed that the QFL approaches achieve performance comparable to a centralized approach while preserving data privacy.
\balance
\bibliographystyle{IEEEtran} 
\bibliography{ref}

\end{document}